\documentclass[10pt,conference]{IEEEtran}

\usepackage[letterpaper, left=0.67in, right=0.67in, bottom=1.02in, top=0.725in]{geometry}

\usepackage{cite}
\usepackage{graphicx}
\usepackage{subcaption}
\usepackage{float} 
\usepackage{mathtools,amssymb,lipsum,caption}
\usepackage{amsmath}
\usepackage[ruled,linesnumbered]{algorithm2e}
\usepackage{epstopdf}
\usepackage{pdfpages}
\usepackage{float}
\usepackage{multirow}
\usepackage{cases}

\usepackage{acronym}
\def\BibTeX{{\rm B\kern-.05em{\sc i\kern-.025em b}\kern-.08em
    T\kern-.1667em\lower.7ex\hbox{E}\kern-.125emX}}

\acrodef{6G}{sixth-generation}
\acrodef{OFDM}{orthogonal frequency division multiplexing}
\acrodef{CSI}{channel state information}
\acrodef{AWGN}{additive white gaussian noise}
\acrodef{ISAC}{Integrated sensing and communication}
\acrodef{BER}{bit error rate}
\acrodef{MSE}{mean square error}
\acrodef{MIMO}{multiple-input multiple-output}
\acrodef{RSS}{received signal strength}
\acrodef{AN}{anchor node}
\acrodef{AP}{access point}
\acrodef{APs}{access points}

\begin{document}

\title{Simultaneous Intrusion Detection and Localization Using ISAC Network }

\author{
\IEEEauthorblockN{Usama Shakoor$^1$, Muhammad Bilal Janjua$^2$, Muhammad Sohaib J. Solaija$^3$, and H\"{u}seyin Arslan$^1$}\\ 
\IEEEauthorblockA{$^{1}$Department of Electrical and Electronics Engineering, Istanbul Medipol University, Istanbul, 34810 Turkey\\
$^{2}$ R\&D Department, Oredata, Istanbul, 34220 Turkey\\
$^{3}$Institute of Defence Technologies Gebze Technical University, Gebze, Kocaeli, 41400 Turkey\\
Email: usama.shakoor*@std.medipol.edu.tr, bilal.janjua@oredata.com, solaija@ieee.org, huseyinarslan@medipol.edu.tr}}

\maketitle

\begin{abstract}
  The rapid increase in utilization of smart home technologies has introduced new paradigms to ensure the security and privacy of inhabitants. In this study, we propose a novel approach to detect and localize physical intrusions in indoor environments. The proposed method leverages signals from access points (APs) and an anchor node (AN) to achieve accurate intrusion detection and localization. We evaluate its performance through simulations under different intruder scenarios. The proposed method achieved a high accuracy of 92\% for both intrusion detection and localization. Our simulations demonstrated a low false positive rate of less than 5\% and a false negative rate of around 3\%, highlighting the reliability of our approach in identifying security threats while minimizing unnecessary alerts. This performance underscores the effectiveness of integrating Wi-Fi sensing with advanced signal processing techniques for enhanced smart home security.
\end{abstract}


\section{INTRODUCTION} \ac{ISAC} systems are increasingly utilized for applications like environmental monitoring and smart infrastructure management \cite{liu2022integrated}. By integrating sensing and communication, \ac{ISAC} enables devices to achieve real-time environmental awareness, beneficial for tasks such as intrusion detection and localization \cite{zou2016survey}. The IEEE 802.11bf standard, for instance, enhances Wi-Fi sensing capabilities, allowing devices to detect movement and localize objects in their surroundings \cite{chen2022wi}. This research leverages these advancements to propose a method for simultaneous intrusion detection and localization using existing Wi-Fi infrastructure, beam sweeping, and \ac{RSS}-based localization for smart home security.

Several methods for intrusion detection and localization in wireless environments have been proposed, including anomaly detection, signature identification, and deep learning \cite{jin2021intrusion, yang2019combined, wang2022collaborative}. Indoor localization techniques such as Wi-Fi fingerprinting and radio frequency (RF) sensor networks have also been studied \cite{mao2007wireless, he2015wi, saeed2019state}, with machine learning and ultra-wideband (UWB) signals enhancing these systems \cite{patwari2010rf}. Simultaneous intrusion detection and localization have been explored using methods like generalized likelihood ratio test (GLRT) and artificial neural networks (ANNs) \cite{alroomi2018secure, gebremariam2023localization}. However, none specifically addresses the simultaneous detection and localization of physical intrusions in indoor environments, crucial for timely and accurate responses in real-time security systems.

This paper proposes a method that integrates intrusion detection and localization by analyzing \ac{RSS} data from multiple \acp{AP}. Unlike traditional methods focusing on either task, our approach uses both signal variation and fluctuations to detect intrusions. By leveraging transmitted signal features at the localized \ac{AN}, the system detects intruders through \ac{RSS} value analysis, reducing reliance on triangulation alone. Utilizing existing Wi-Fi infrastructure, our approach offers a cost-effective and scalable solution without requiring additional hardware. Beam sweeping enhances detection precision by capturing fine-grained \ac{RSS} values in multiple directions. Our simulations show an accuracy of over 92\% for both intrusion detection and localization, with a low false positive rate of less than 5\% and a false negative rate of around 3\%, surpassing many existing techniques.

Our main contributions to the \ac{ISAC} literature are as follows: \begin{itemize} \item Introduction of a triangulation-based approach to localize the \ac{AN} and perform coarse intrusion detection by comparing \ac{RSS} values over regular intervals. \item Implementation of beam sweeping for fine intrusion detection, capturing \ac{RSS} values in all directions to pinpoint the intruder’s presence at specific angles. \item Demonstrating significant performance gains, achieving 92\% accuracy in detecting and localizing physical intrusions, with a false positive rate of less than 5\% and false negatives around 3\%. \end{itemize}

\begin{figure*}[t] 
    \centering
    \begin{subfigure}[b]{0.41\textwidth} 
        \includegraphics[width=\linewidth]{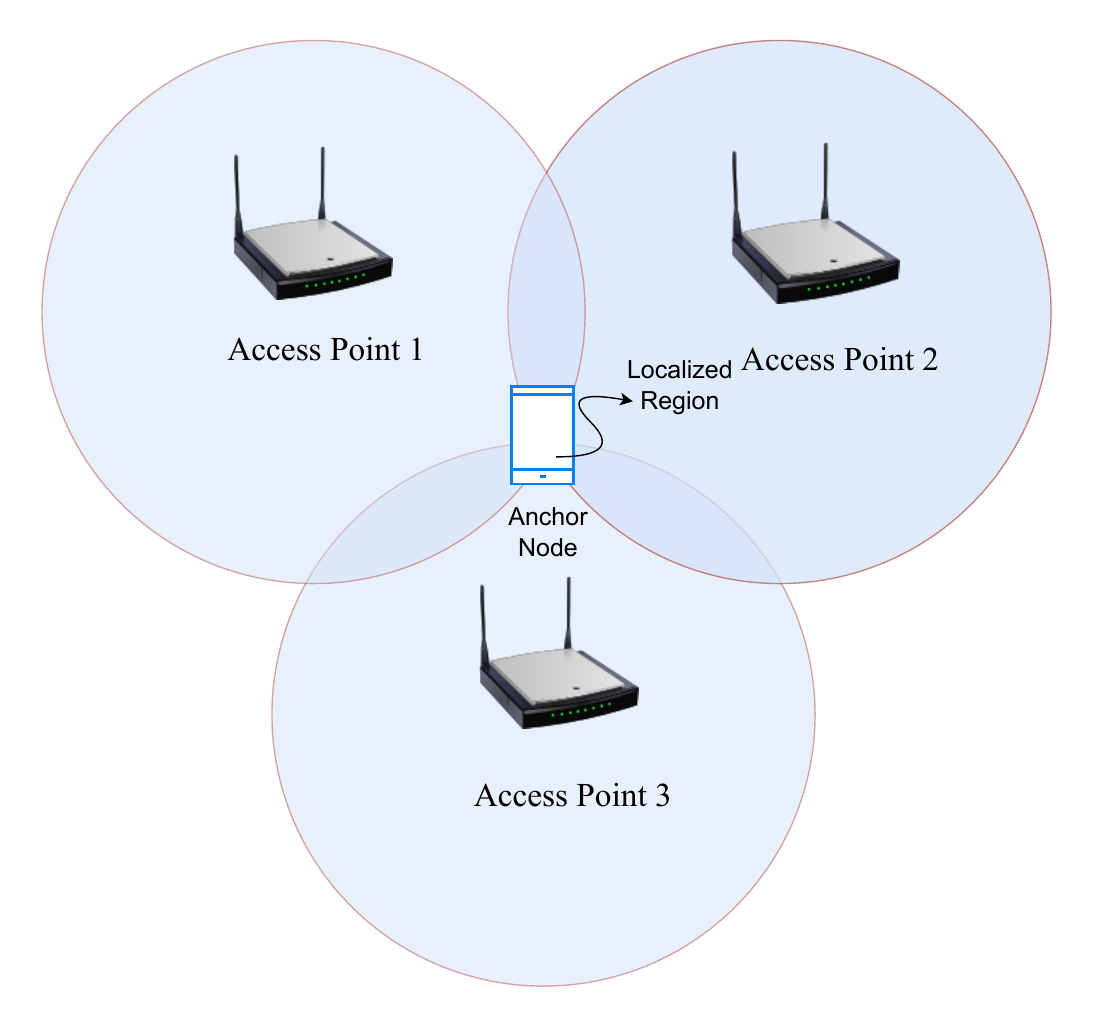}
        \caption{Triangulation setup for observing reference \ac{RSS} values at the \ac{AN} from multiple \acp{AP}.}
        \label{1}
    \end{subfigure}
    \hfill
    \begin{subfigure}[b]{0.41\textwidth} 
        \includegraphics[width=\linewidth]{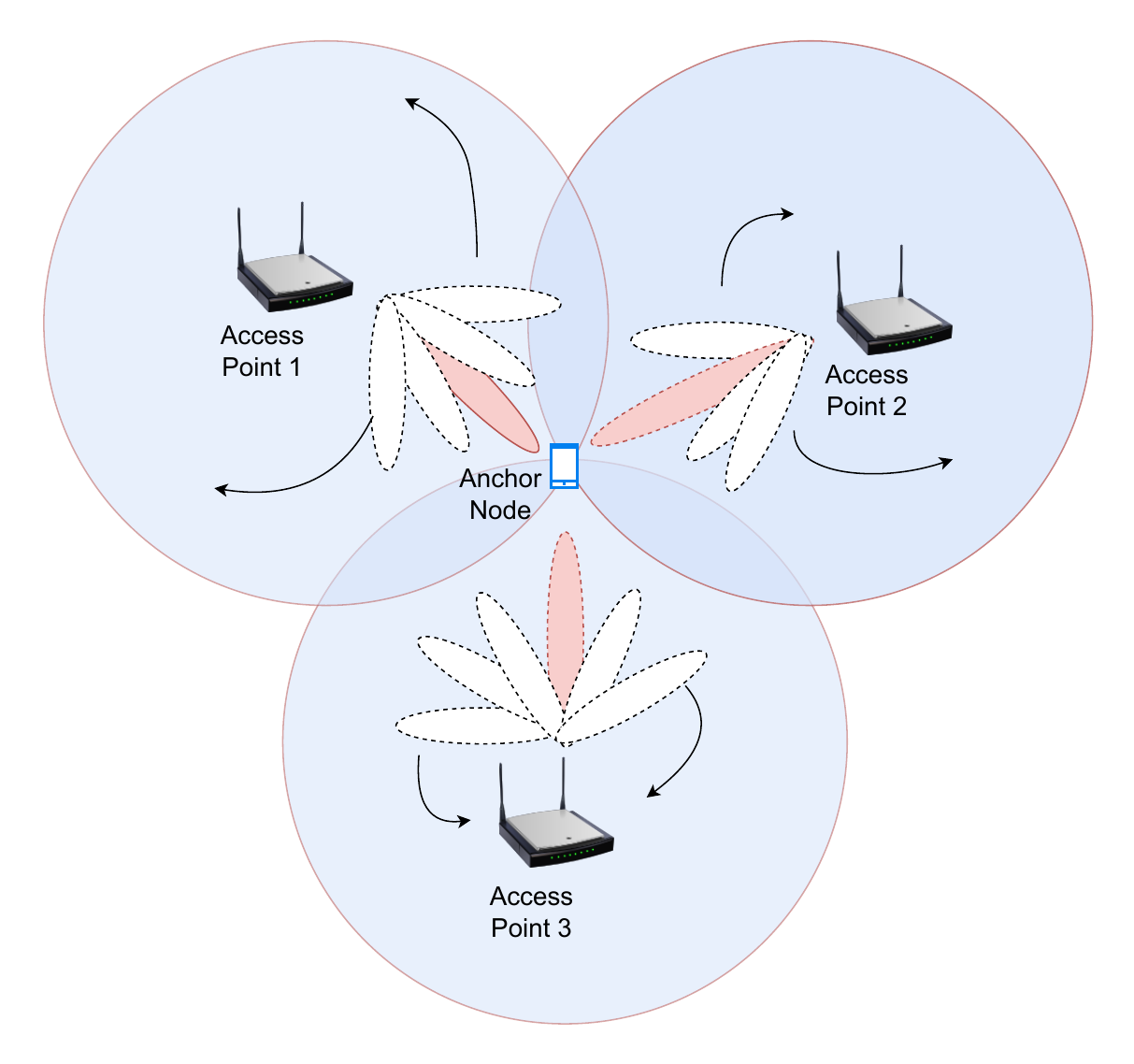}
        \caption{Beam sweeping for recording \ac{RSS} values at different angles at the \ac{AN}.}
        \label{2}
    \end{subfigure}
    \hfill
    \begin{subfigure}[b]{0.41\textwidth} 
        \includegraphics[width=\linewidth]{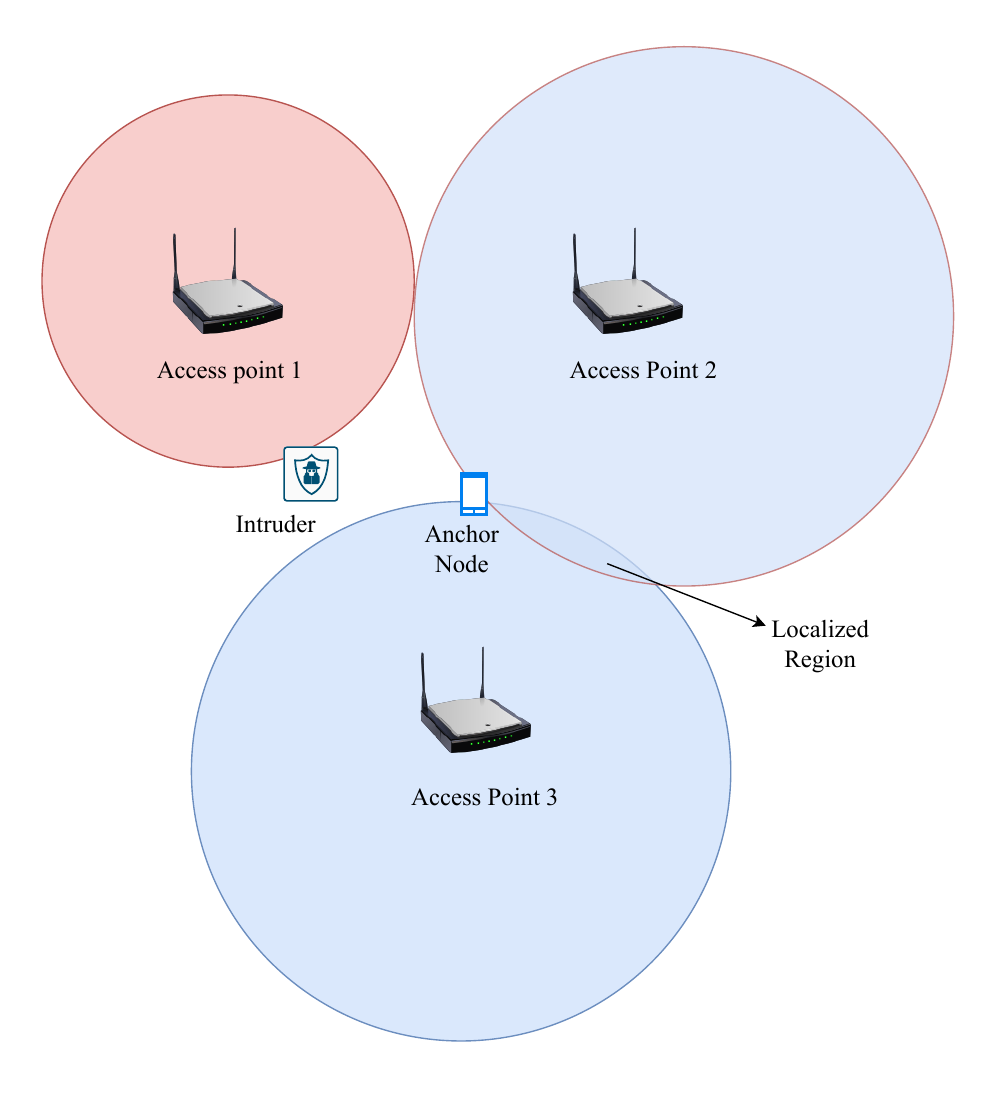}
        \caption{Intrusion detection and coarse localization: \ac{RSS} of one of the \acp{AP} reduces significantly in the presence of an intruder observed at the \ac{AN}.}
        \label{3}
    \end{subfigure}
    \hfill
    \begin{subfigure}[b]{0.41\textwidth} 
        \includegraphics[width=\linewidth]{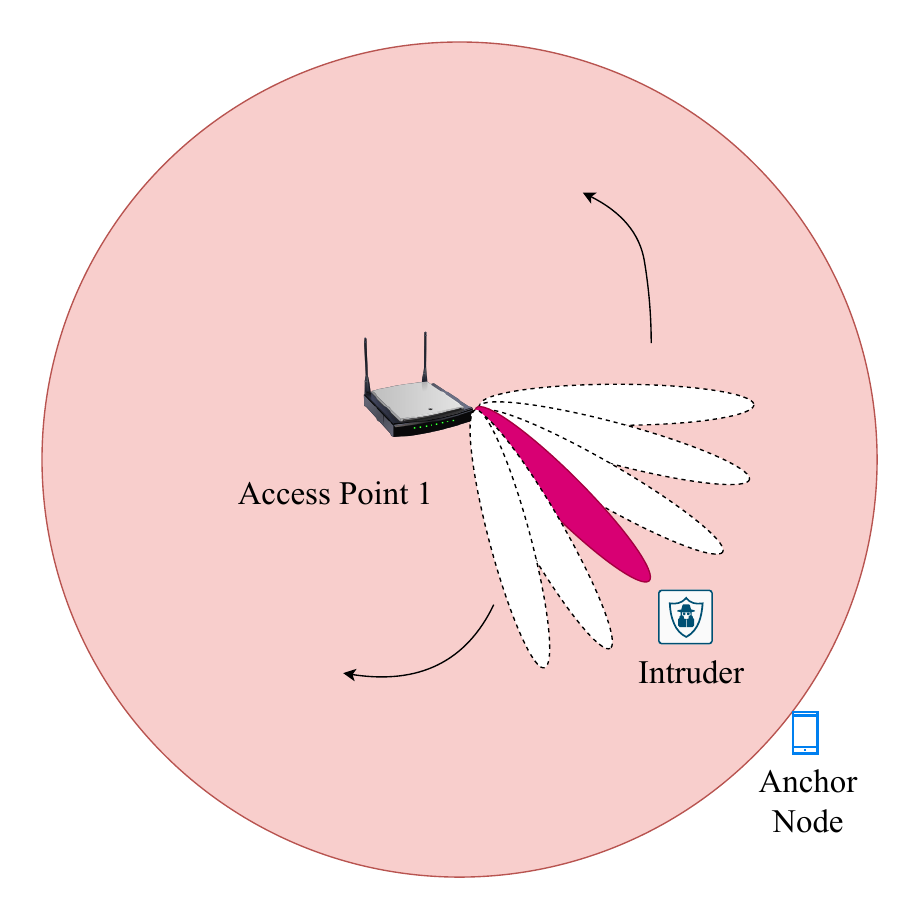}
        \caption{Fine intrusion detection by beam sweeping again and observing fluctuations in \ac{RSS} compared with the previous step.}
        \label{4}
    \end{subfigure}
    \caption{ Illustration of intrusion detection and localization process in a smart home environment using RSS and beam sweeping techniques.}
    \label{fig:combined_figures_1_to_4}
\end{figure*}

\section{System Model}
The system comprises of $K$ \acp{AP}, where in this work, we assume three \acp{AP} $k_1$, $k_2$, and $ k_3$. Each \ac{AP} act as transmitters, each equipped with a uniform rectangular array (URA) of $M \times N$ \ac{MIMO} antennas. Here, we assume that the multiple antennas at both the \acp{AP} and \ac{AN} improve signal diversity, spatial multiplexing, and reliability, with \ac{MIMO} supporting both downlink and uplink communications between the \acp{AP} and the users or \ac{AN}. The presence of multiple antennas at the \ac{AN} enhances the system's ability to accurately capture signals from different directions, thereby improving localization accuracy.

An \ac{AN} $a$ exists within the environment, strategically positioned to measure the RSS values between each \ac{AP} and the \ac{AN} pair $(a, k_i)$, where $k_i \in K$. The \ac{AN} may also have multiple antennas, benefiting from the MIMO configuration to capture more diverse signal paths and improve detection accuracy.

Each \ac{AP} transmits an \ac{OFDM} signal to the \ac{AN}, and the RSS values are measured from each \ac{AP}-\ac{AN} pair. \ac{OFDM} is chosen for its ability to mitigate intersymbol interference (ISI) and handle multipath propagation, which is crucial in indoor environments. While IFFT is applied to convert the complex data symbols from the frequency domain, $X_c = [X_c(0), X_c(1), \dots, X_c(N-1)]^T$, to the time domain symbols $x_c = [x_c(0), x_c(1), \dots, x_c(N-1)]^T$. This step is necessary to create an \ac{OFDM} waveform for transmission, rather than for specific time-domain processing. 

The IFFT process is crucial in \ac{OFDM} for modulating the subcarriers, ensuring orthogonality in the transmitted signal. The time-domain representation is used primarily for transmission over the wireless channel, where the \ac{AN} measures the signal properties. Time-domain processing is not performed explicitly after this step, but the conversion allows for more efficient transmission and analysis of multipath effects, shadowing, and fading. The IFFT operation is performed as follows:
\begin{equation} \label{Eq:03}
    x_c(t)= \sum_{k=0}^{N-1} X_c(k) e^{j 2 \pi k \Delta f t} \quad 0 \leq t \leq T_s~.
\end{equation}
Then, the sampled version of the \ac{OFDM} symbol is denoted as
\begin{equation}\label{Eq:04}
    x_c(n)=\frac{1}{\sqrt{N}}\sum_{k=0}^{N-1}X_c(k)e^{j2\pi nk/N}.
\end{equation}

\acp{AP} are strategically distributed throughout the smart home environment, ensuring comprehensive coverage and minimizing blind spots. Additionally, the \ac{AN} is strategically placed to serve as a reference point for localization purposes. The \ac{AN}'s placement can be predefined or dynamically adjusted to optimize localization accuracy. Intruders within the environment are characterized by their presence in the vicinity of one or more \acp{AP}, potentially causing fluctuations or reductions in the \ac{RSS}.

The signal model considers factors such as multipath fading, shadowing, and other propagation effects, influencing the observed \ac{RSS} values at the \ac{AN}. Let $r_{a,k}(i)$ be the \ac{RSS} measured at the \ac{AN} for transmission from \ac{AP} $k$ at time $i$. The \ac{RSS} vector is defined as:
\begin{equation}
    \mathbf{r}_a (i)=\left [ r_{a,k_0}(i),,...,r_{a,k_n}(i) \right ]^T,
\end{equation}
and the mean of the \ac{RSS} vector over a window time $T$ is:
\begin{equation}
\mathbf{\bar{r}}_a (i)=\frac{1}{T}\sum_{t=1}^{T}\mathbf{r}_a (i-t).
\end{equation}

\section{Proposed Method}
This section delineates the systematic approach employed to realize the objectives of our study. We start by discussing the initial step of establishing a spatial reference point, which, while not always mandatory, is crucial in our methodology for accurately performing subsequent intrusion detection and localization tasks. Following this, we delve into synthetic \ac{RSS} data generation, where simulated datasets are meticulously crafted to emulate real-world scenarios, facilitating the evaluation of our proposed methods. Subsequently, we explore intrusion detection, detailing the algorithms and techniques utilized to identify anomalous behavior indicative of potential intrusions. Finally, we discuss intrusion localization, where the precise location of detected intrusions is determined utilizing location-based methodologies, enhancing the security and situational awareness of smart home environments. A detailed description of the procedures followed is presented in Algorithm 1.

\subsection{\ac{AN} Localization}

In this step, our objective is to accurately determine the position of the \ac{AN} within the smart home environment. The precise localization of the \ac{AN} is essential as it serves as the reference point for subsequent intrusion detection and localization tasks. The \ac{AN}’s position is estimated using a triangulation technique that leverages \ac{RSS} measurements from multiple \acp{AP} distributed throughout the environment, as illustrated in Fig. \ref{1}.

To estimate the \ac{AN}’s position, we first compute the distances between the \ac{AN} and each \ac{AP} using the observed \ac{RSS} values. These distances are then used in conjunction with the known positions of the \acp{AP} to triangulate the position of the \ac{AN}. The fundamental equations involved in this process are:

\begin{equation} \operatorname{RSS}_{\text {BL}, i}=P_t - 20 \log{10} \frac{4 \pi d_i}{\lambda} + \sigma_n^2, \end{equation}

\begin{equation} d_i = \sqrt{\left(x_a - x_i\right)^2 + \left(y_a - y_i\right)^2}, \end{equation}
where $\mathrm{RSS}_{\text{BL}, i}$ represents the baseline \ac{RSS} observed at the \ac{AN} from \ac{AP} $k_i$. $P_t$ denotes the transmit power of the \ac{AP}, $\sigma_n^2$ is the noise power, and $\lambda$ is the signal wavelength. $d_i$ is the distance between the \ac{AN} and \ac{AP} $k_i$. $\left(x_a, y_a\right)$ and $\left(x_i, y_i\right)$ are the coordinates of the \ac{AN} and  \ac{AP} $k_i$, respectively.

The \ac{RSS} observed at the \ac{AN} from each \ac{AP} is calculated as the sum of the transmit power, the path loss component, and the noise power. The path loss component is determined using the Friis transmission equation \cite{franek2017phasor}, which accounts for the distance between the \ac{AN} and each \ac{AP}, as well as the wavelength of the transmitted signal. By measuring the \ac{RSS} from multiple \acp{AP} and knowing their positions, we can accurately triangulate the \ac{AN}’s position within the smart home environment. This triangulation provides a baseline reference for subsequent localization calculations, which is essential for detecting and localizing intrusions effectively.

\subsection{Synthetic RSS Data Generation}

In this step, we generate synthetic RSS data to simulate measurements obtained from the \acp{AP} under different scenarios. Figure \ref{2} illustrates the process of generating synthetic RSS data, which includes beam sweeping to capture RSS values at various angles.
The synthetic RSS data is generated for two distinct scenarios: (1) normal scenario (without intrusion) and (2) scenario with intrusion.

\subsubsection{Scenario Without Intrusion}

In the normal scenario, the RSS value is calculated for each sample $t$, \ac{AP} $k_i$, and angle $j$. This value represents the RSS measurement without any intrusion present. The formula for calculating the RSS value is given by:

\begin{equation}
R_{\text {No\_intr}, t, i, j}=\text{RSS}_{\text {BL}, i}+\mathcal{F}+ \mathcal{S}.
\end{equation}
where
$\text{RSS}_{\text{BL}, i}$ is the baseline RSS observed at the \ac{AN} from \ac{AP} $k_i$. $\mathcal{F}\sim \mathcal{N}(0, \sigma_f^2)$ represents  random variation in signal strength due to multipath fading effects, modeled as a zero-mean Gaussian random variable with standard deviation $\sigma_f$. $\mathcal{S}\sim \mathcal{N}(0, \sigma_s^2)$ represents random variation in signal strength due to obstacles, modeled as a zero-mean Gaussian random variable with standard deviation $\sigma_s$.


\subsubsection{Scenario with Intrusion}

In the presence of an intrusion, the RSS value is reduced due to the obstruction caused by the intruder. We model the reduction in RSS as a gradual change based on the distance between the \ac{AN} and the intruder. The intrusion-affected RSS value is modeled as
%
\begin{equation}
\text{RSS}_{\text{intr}, t, i, j} = 
\begin{cases}
\text{RSS}_{\text{BL}, i} - \Delta \text{RSS} + \mathcal{F} \\ + \mathcal{S}, 
\quad ||(x_a, y_a) - (x_i, y_i)|| < \tau_d \\
\text{RSS}_{\text{BL}, i} + \mathcal{F} + \mathcal{S}, \quad \text{otherwise}
\end{cases}.
\label{Eq:10}
\end{equation}
%
where $\Delta \text{RSS}$ represents the average RSS reduction due to an intruder, which can vary depending on the specific characteristics of the intrusion, $\tau_d$ is the distance threshold at which the effect of the intruder becomes significant. This is typically adjusted to reflect realistic conditions. 

\subsection{Intrusion Detection} 
For each sample $t$ in the dataset, the system analyzes the \ac{RSS} data obtained from the \acp{AP}. It computes the deviation  of \ac{RSS} values $\mathcal{D}$ from their respective means for each \ac{AP}. A high value of $\mathcal{D}$  indicates a significant change in signal strength, potentially caused by the presence of an intruder. The maximum $\mathcal{D}$ across all \acp{AP} is then compared against the predefined threshold $\tau$. If the maximum $\mathcal{D}$ exceeds $\tau$, intrusion detection is triggered for that sample. Figure 1c presents the case of intrusion detection. The mathematical representation for intrusion detection is given by 
\begin{equation}
\text{intr detected} = \begin{cases}
\text{true}, & \max(\mathcal{D}) > \tau \\
\text{false}, & \text{otherwise}
\end{cases}~, \\
\end{equation}
and
\begin{equation}
    \mathcal{D} = \left| \text{RSS}_{\text{intr}, t, i, j} - \mathbb{E}\left[\text{RSS}_{\text{intr}, t, i, :}\right] \right|~,
\end{equation}
where $\mathbb{E[\cdot]}$ represents the expected value. The value of $\tau$ is crucial in determining the sensitivity of the intrusion detection algorithm. A higher $\tau$ may lead to fewer false alarms but could potentially miss some intrusions, while a lower $\tau$ may result in more false alarms but higher detection sensitivity. The $\tau$ can be adjusted based on the specific requirements and constraints of the smart home environment, balancing between detection accuracy and false alarm rate.

\subsection{Intrusion Localization} 
The intruder's location is estimated by calculating the angle and distance relative to the AN as illustrated in Fig. 1d. The angle $\theta_{\text{intr}}$ is determined from the beam sweeping procedure, while the distance is derived from the reduction in RSS. The estimated coordinates of the intruder are given by:

\begin{equation}
\begin{aligned}
(x_{\text{intr}}, y_{\text{intr}}) =\,& (x_a, y_a) \\
&+ d_{\text{intr}} \cdot (\cos(\theta_{\text{intr}})~, \sin(\theta_{\text{intr}}))~,
\end{aligned}
\end{equation}
where $d_{\text{intr}}$ is the estimated distance of the intruder from the AN, based on the degree of RSS attenuation. $\theta_{\text{intr}}$ is the angle of intrusion, determined from beam sweeping at the AP. $(x_a, y_a)$ are the known coordinates of the AN. The distance $d_{\text{intr}}$ is calculated by using the attenuation model of RSS, where a significant drop in signal strength corresponds to the proximity of the intruder.

\begin{algorithm}
  \SetAlgoLined
  \KwData{\ac{AN} position $(x_a, y_a)$, \acp{AP} positions $(x_i, y_i)$, \ac{RSS} baseline $RSS_{\text{BL}}$, distance threshold $\tau_d$, and detection threshold $T$}
  \KwResult{Intrusion detection status $I_\mathrm{D}$ and localization $L$}
  
  Generate synthetic \ac{RSS} data for normal scenario ($RSS_{\text{No\_intr}}$) and intrusion scenario ($RSS_{\text{intr}}$).
  Initialize intrusion detection status as $I_\mathrm{D} = \text{false}$
  Initialize intrusion location as unknown $L = \emptyset$.
  \ForEach{data point $I_\mathrm{D}$ in $RSS_{\text{No\_intr}}$}{
    Perform intrusion detection: $I_\mathrm{D} \leftarrow \text{Detect intrusion}(d, T)$\\
    \If{$I_\mathrm{D}$ is true}{
      \ForEach{data point $I_\mathrm{D}$ in $RSS_{\text{intr}}$}{
        Perform fine intrusion detection by analyzing $\mathcal{D}$\\
        \If{intrusion detected}{
          Perform intrusion localization: \\$L \leftarrow$ Localize intrusion $(d, (x_a, y_a), \tau_d)$\\
          \textbf{Break}
        }
      }
      \textbf{Break}
    }
  }  
  \caption{Intrusion detection and localization algorithm}
\end{algorithm}

\begin{table}[!ht]
    \centering
    \caption{Simulation Parameters}
    \begin{tabular}{|l|l|}
        \hline
        \textbf{Parameter} & \textbf{Value} \\ \hline
        Number of \acp{AP} & 3 \\ \hline
        Resolution of Beam Sweeping (degrees) & 360 \\ \hline
        Intrusion Angle (degrees) & 120 \\ \hline
        Path Loss Exponent & 3 \\ \hline
        Number of OFDM Subcarriers & 64 \\ \hline
        Intrusion Effect (dB) & -10 \\ \hline
        Distance Between Anchor and APs (meters) & 5 \\ \hline
        Number of Simulations & 100 \\ \hline
        Detection Threshold (degrees) & 20 \\ \hline
        \end{tabular}
    \label{tab:simulation_parameters}
\end{table}
 
\section{Results and Discussions}

In this section, we present the results obtained from our simulations, focusing on the \ac{RSS} signals from \acp{AP} in the presence and absence of intrusions, the system performance metrics, and the root mean square error (RMSE) of intrusion localization.
Table \ref{tab:simulation_parameters} summarizes the key parameters used in our simulations.

\begin{figure}[ht]
    \centering
    \includegraphics[width=\linewidth]{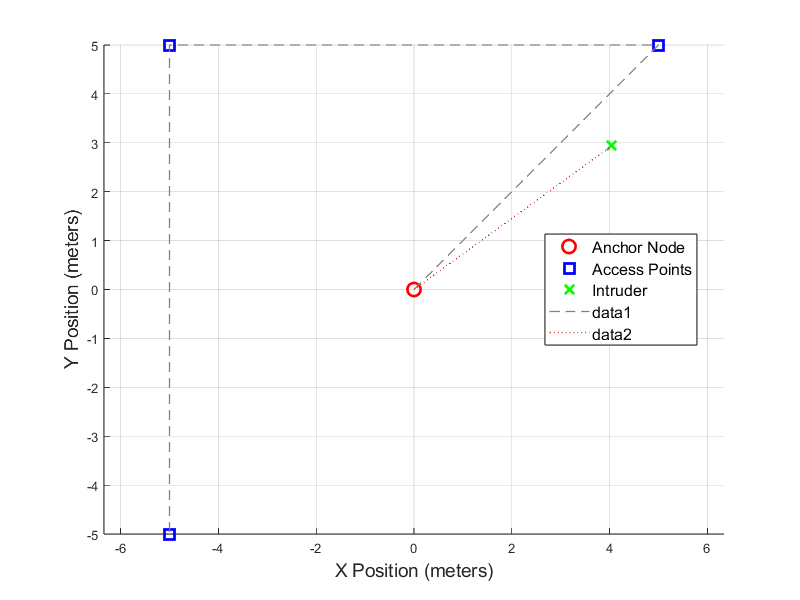}
    \caption{\ac{AN}, \acp{AP}, and Intruder Location.}
    \label{fig:node_locations}
\end{figure}

The placement of the \ac{AN}, \acp{AP}, and the intruder is depicted in Fig. \ref{fig:node_locations}. This provides a visual representation of the spatial arrangement during the intrusion detection process. Data1 refers to the baseline RSS values collected from the \acp{AP} without any intrusion effect, whereas, data2 represents the RSS values with the intrusion effect applied. These values include a reduction in RSS at specific angles due to the presence of an intruder, allowing for comparison between normal and altered RSS readings.

\begin{figure}[ht]
    \centering
    \includegraphics[width=\linewidth]{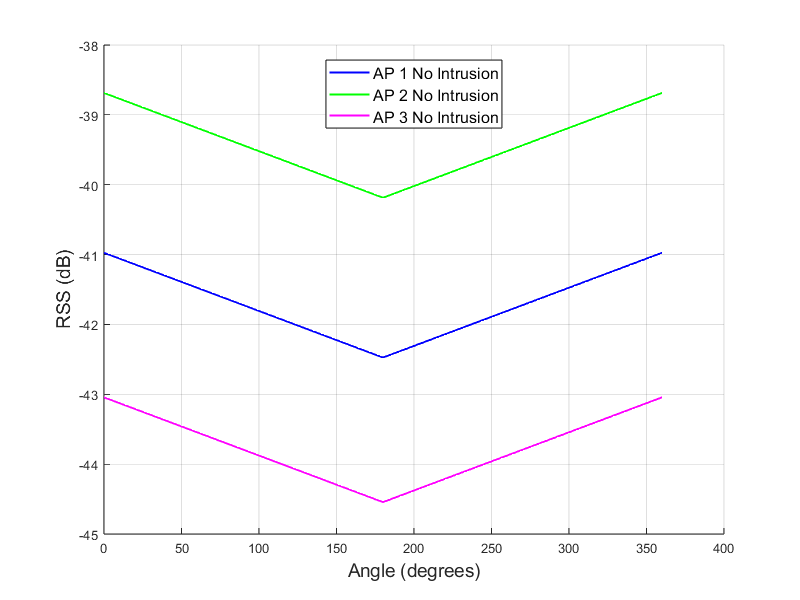}
    \caption{Baseline RSS signals from \acp{AP}.}
    \label{fig:baseline_rss}
\end{figure}


Figure \ref{fig:baseline_rss} illustrates the baseline RSS (dB) signals for different angle in degrees from the three \acp{AP} without any intrusion. Each \ac{AP} exhibits distinct \ac{RSS} values, demonstrating the effects of distance and angle variation. For example, the RSS of \ac{AP} 1 is above $-39$dB at $0$ degree and it degrees with the increasing angle until $200$ degrees and then it starts to increases again. \ac{AP} 2 and 3 shows the same behavior but with different RSS values.

\begin{figure}[!ht]
    \centering
    \includegraphics[width=\linewidth]{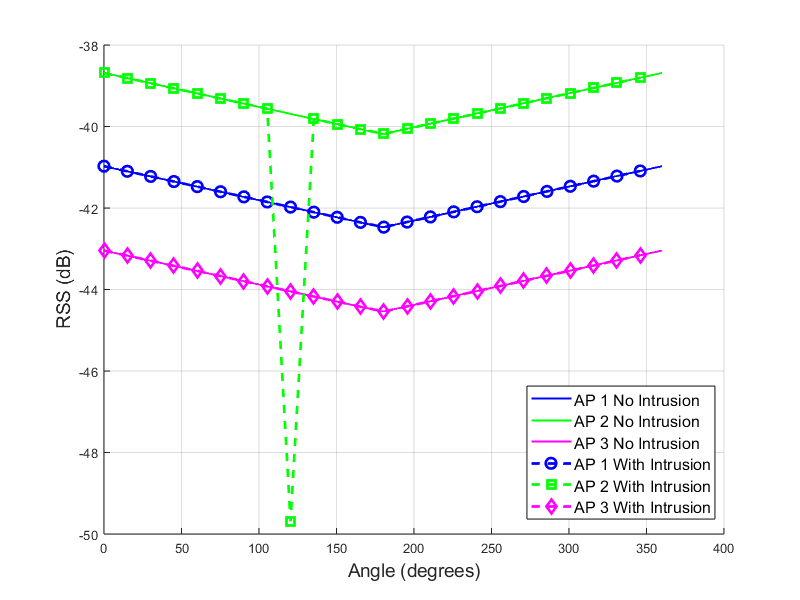}
    \caption{RSS signals with intrusion effect.}
    \label{fig:intrusion_rss}
\end{figure}

\begin{figure}[ht]
    \centering
    \includegraphics[width=\linewidth]{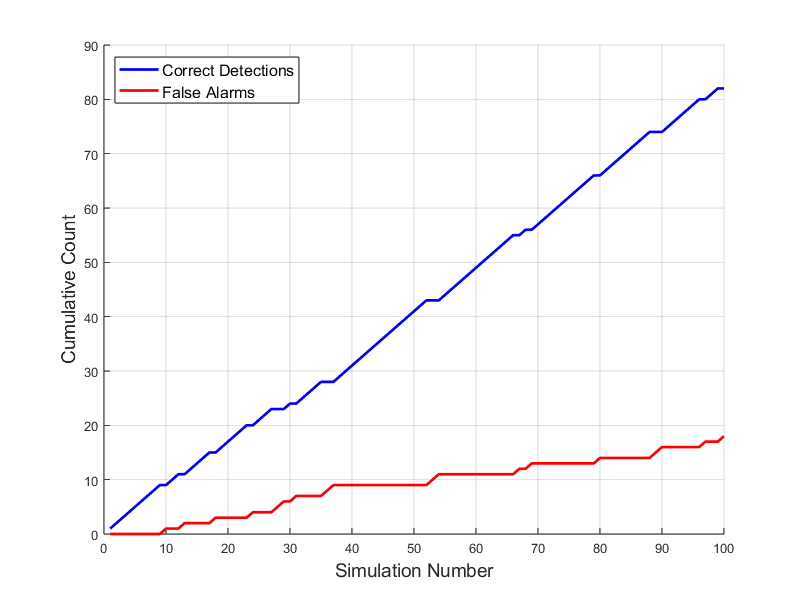}
    \caption{System's detection performance for different simulation numbers.}
    \label{fig:performance_detection}
\end{figure}

Figure \ref{fig:intrusion_rss} shows the RSS signals when an intrusion occurs. The intrusion effect is observed primarily in the RSS from AP$_2$, demonstrating the impact of the intruder on the RSS measurements.
%
%
The cumulative counts of correct detections and false alarms over the simulations are presented in \figurename~\ref{fig:performance_detection}. The results indicate that our system can effectively identify intrusions with a controlled false alarm rate. For examples, at the simulation count of $50$ the correct detections reach over $40$ at a false alarm below $10$.

\begin{figure}[ht]
    \centering
    \includegraphics[width=\linewidth]{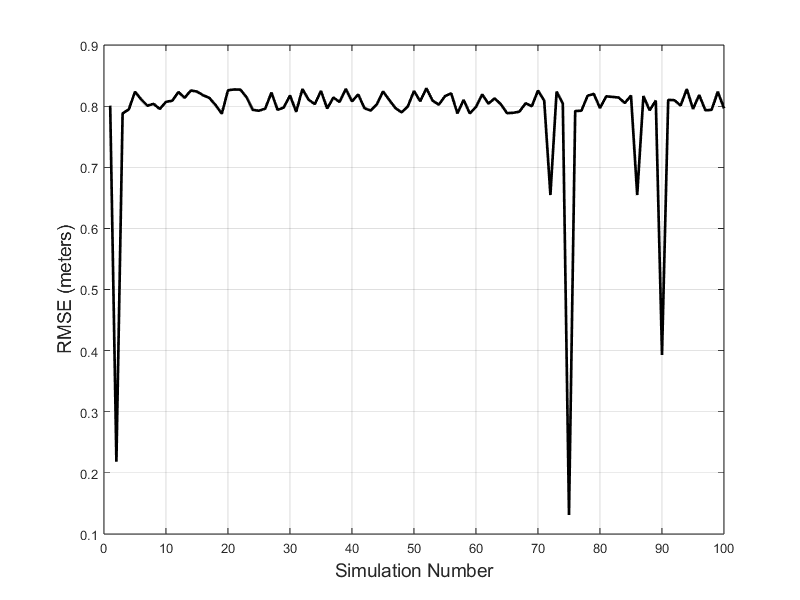}
    \caption{RMSE of intruder localization different simulation numbers.}
    \label{fig:rmse_localization}
\end{figure}

The RMSE of the estimated intrusion angle, converted to meters, is depicted in \figurename~\ref{fig:rmse_localization}. This metric provides insight into the localization accuracy of our method, highlighting the effectiveness of the proposed approach. 
Simulation results demonstrate the proposed method's effectiveness for simultaneous intrusion detection and localization using the ISAC approach. The distinct RSS patterns and low RMSE values illustrate the feasibility of our approach in real-world applications.

\section{Conclusion}

In this paper, we proposed and evaluated a novel method for detecting and localizing physical intrusions in smart homes using an ISAC system. Our approach leverages multiple \acp{AP} and an \ac{AN} to detect changes in the \ac{RSS} caused by intrusions and accurately localize the intruder's position. The simulation results demonstrate the effectiveness of our proposed method. The baseline RSS measurements showed expected variations with beam sweeping, and the introduction of an intrusion effect resulted in significant changes in the RSS values, which were successfully detected by our system. Our system achieved high accuracy in intrusion detection, with correct detection rates improving consistently across simulations. The false alarm rate was kept low, indicating robust performance. We also evaluated the localization accuracy of our system, which exhibited a low RMSE, reflecting the precision of our method in determining the intruder's position. The visual representation of node locations and the system performance metrics further validate the effectiveness and practicality of our approach. Overall, our results highlight the potential of using ISAC techniques to enhance security in smart homes. Future work could focus on integrating additional sensors or refining the algorithm to handle more complex scenarios, such as varying environmental conditions or multiple simultaneous intrusions. Further exploration into real-world implementations and testing could provide additional insights into the scalability and robustness of the proposed method.

\section*{Acknowledgment}
This work is supported by The Scientific and Technological Research Council of Türkiye (TÜBİTAK) 1515 Frontier R\&D Laboratories Support Program for Turk Telekom neXt Generation Technologies Lab (XGeNTT) under project number 5249902.

\addtolength{\textheight}{-12cm}  
\bibliographystyle{IEEEtran}
\bibliography{references}

\begin{IEEEbiography}

\end{IEEEbiography}

\end{document}